\documentclass[runningheads]{llncs}
\usepackage{graphicx}
\usepackage{amsmath}
\usepackage{physics}
\usepackage{amssymb}
\usepackage{float}
\usepackage{subcaption}
\usepackage{algpseudocode}
\usepackage{algorithm}
\usepackage[]{mdframed}
\usepackage{xcolor}

\newcommand\blfootnote[1]{%
  \begingroup
  \renewcommand\thefootnote{}\footnote{#1}%
  \addtocounter{footnote}{-1}%
  \endgroup
}

\begin{document}
\title{Feedback-Based Quantum Algorithm for Constrained Optimization Problems}
\titlerunning{Feedback-Based Quantum Algorithm for Constrained Optimization Problems}
%
\author{Salahuddin Abdul Rahman\inst{1}\orcidID{0009-0002-9686-8586}  \and
Özkan Karabacak\inst{2}\orcidID{0000-0002-8350-193X} \and
Rafal Wisniewski \inst{1}\orcidID{0000-0001-6719-8427}}
\authorrunning{S. Abdul Rahman et al.}
%
\institute{Automation and Control section, Department of electronic systems, Aalborg University, Aalborg, Denmark \\ \email{\{saabra, raf\}@es.aau.dk} \and  
Department of Mechatronics Engineering, Kadir Has University,  Istanbul, Turkey
\email{ozkan.karabacak@khas.edu.tr}}
\maketitle              
\begin{abstract}

The feedback-based algorithm for quantum optimization \\(FALQON) has recently been proposed to find ground states of Hamiltonians and solve quadratic unconstrained binary optimization problems. This paper efficiently generalizes FALQON to tackle quadratic constrained binary optimization (QCBO) problems. For this purpose, we introduce a new operator that encodes the problem's solution as its ground state. Using control theory, we design a quantum control system such that the state converges to the ground state of this operator. When applied to the QCBO problem, we show that our proposed algorithm saves computational resources by reducing the depth of the quantum circuit and can perform better than FALQON. The effectiveness of our proposed algorithm is further illustrated through numerical simulations.

\blfootnote{This paper is accepted for publication in the 15th International Conference on Parallel Processing and Applied Mathematics (PPAM 2024).}
\end{abstract}
%
%
%

\section{Introduction} \label{s1}
For noisy intermediate-scale quantum (NISQ) computers, the leading algorithms that can fulfil these devices' requirements and are expected to show quantum advantage are the variational quantum algorithms (VQAs) \cite{bharti2022noisy}.
VQAs have applications in quantum chemistry, error correction, quantum machine learning, and combinatorial optimization \cite{cerezo2021variational}. Two primary challenges VQAs face are the design of the ansatz and the need to solve a challenging classical optimization problem to update the parameters of the parameterized quantum circuit \cite{cerezo2021variational}. To overcome these challenges, several techniques were proposed, such as layer-wise construction of the ansatz, for example, in adaptive derivative-assembled pseudo-trotter variational quantum eigensolver (ADAPT-VQE) \cite{grimsley2019adaptive}, and by using suitable classical optimizer \cite{fernandez2022study}.  



In \cite{magann2022lyapunov,magann2022feedback}, Magann et al. proposed the feedback-based algorithm for quantum optimization (FALQON) algorithm as an alternative approach for solving quadratic unconstrained binary optimization (QUBO) problems. FALQON constructs the quantum circuit layer-by-layer and assigns the circuit parameters through measurements of the qubits from the previous layer. This algorithm has the advantage of avoiding using a classical optimizer and resulting in a monotonic improvement of the approximate solution with the increasing depth of the circuit. In \cite{larsen2023feedback}, the feedback-based quantum algorithm (FQA) was proposed to extend FALQON to prepare the ground states of Hamiltonians, specifically in the Fermi-Hubbard model. In addition, the feedback-based quantum algorithm for excited states calculation (FQAE) \cite{rahman2024feedback} and the weighted feedback-based quantum algorithm for excited states calculation (WFQAE) \cite{rahman2024weighted} were proposed to extend feedback-based quantum algorithms to prepare excited states of Hamiltonians. Despite the potential of FALQON, it usually results in deeper circuits than the quantum approximate optimization algorithm (QAOA). However, as shown in \cite{magann2022lyapunov}, FALQON can be used as a potential initialization technique for the parameters of QAOA, drastically increasing its performance. 

FALQON is specifically tailored to tackle QUBO problems. When employing FALQON to address a quadratic constrained binary optimization (QCBO) problem, the QCBO problem must first be transformed into an equivalent QUBO problem. Following this transformation, FALQON can be applied to solve the newly formed QUBO problem. For example, in \cite{DavidWakeham2021}, FALQON is applied to the maximum clique problem, which is modelled as a QCBO problem. The problem is first converted to an equivalent QUBO problem by incorporating the constraints into the cost function using a penalty term. However, this method complicates the implementation of the algorithm since the new equivalent QUBO problem will be transformed into an Ising Hamiltonian, including extra terms that come from the constraints, which increase with the increase in the number of constraints. This work presents a significant enhancement to FALQON for tackling QCBO problems. Instead of converting the QCBO problem into a QUBO problem, we directly tackle the QCBO by introducing a new operator that encodes the optimal feasible solution as its ground state. We design this operator by shifting the energies of all infeasible outcomes such that the ground state of this operator is the minimum-energy feasible solution. Subsequently, we design a Lyapunov control law to converge from all initial states to the ground state of this operator. Based on this control framework, we introduce the feedback-based algorithm for quantum optimization with constraints (FALQON-C). We show that FALQON-C when applied to the QCBO problem, saves computational resources by reducing the depth of the circuit and can perform better than FALQON when applied to the equivalent QUBO problem.

Our design methodology for FALQON-C is based on quantum control theory. Rather than altering the control system design by using a new cost Hamiltonian formed by converting the QUBO problem into an Ising Hamiltonian, we modify the Lyapunov function instead of the system's drift Hamiltonian, informing the system that the targeted eigenstate is changed. The modified Lyapunov function uses a new observable that encodes the information about the new targeted eigenstate. This approach significantly improves the implementation of the quantum circuit of the cost operator by eliminating unnecessary components in the circuit implementation due to the constraints. Only the control law is modified to incorporate the newly constructed operator, thereby updating the system about the change in the targeted eigenstate. This demonstrates the potential of control theory to advance the design of efficient quantum algorithms.

We highlight our main contributions as follows. In this work, we consider a discrete model instead of the continuous Schr\"odinger equation and apply quantum Lyapunov control directly on the circuit level. This discrete model offers the advantage of being directly realizable as quantum gates. For further insights into optimization on the circuit level, refer to \cite{magann2021pulses}. Additionally, we introduce FALQON-C, which efficiently generalizes FALQON to address QCBO problems. We demonstrate its effectiveness in solving QCBO and compare it with the direct utilization of FALQON for QCBO problems. 

The remainder of the paper is structured as follows. Section~\ref{s2} gives an overview of QCBO problems and FALQON. Following this, Section 3 presents our proposed algorithm, namely, FALQON-C. To demonstrate the effectiveness of FALQON-C, we present an illustrative example in Section 4. Finally, Section 5 concludes the paper and outlines avenues for future research.


\newcommand{\realv}[1]{\ensuremath{\mathbb{R}^{#1}}}

\section{Feedback-Based Algorithm for Quantum Optimization} \label{s2}
In this section, we define QUBO and QCBO problems, and introduce FALQON \cite{magann2022feedback,magann2022lyapunov}.

The general QUBO problem is given as follows:
\begin{equation} \label{QUBO}
    \min _{x \in \{0,1\}^n} J(x) =\min _{x \in \{0,1\}^n} x^TQ_Jx+c_J^Tx+a_J
\end{equation}
Here, $Q_J \in\realv{n\times n}$ is a symmetric matrix, $c_J$ $\in\realv{n}$ and $a_J$ $\in \realv{}$. FALQON is originally proposed to tackle QUBO problems \cite{magann2022feedback,magann2022lyapunov}. In this work, we extend it to tackle QCBO problems defined in the following general form:
\begin{subequations}
	\label{QCBO}
	\begin{align}
		&\min _{y \in \{0,1\}^n} F(y) = \min _{y \in \{0,1\}^n} y^TQ_Fy+c_F^Ty+a_F\\
		&\text{s.t.} \quad G^{(j)}(y) = y^TQ_G^{(j)}y+{c_G^{(j)}}^T y+a_G^{(j)} \leq 0, \;\;\; j=\{1, 2,3, ...,k_1\} \\
            & \quad \quad V^{(q)}(y) =  y^TQ_V^{(q)}y+{c_V^{(q)}}^Ty+a_V^{(q)} = 0, \;\;\; q=\{1, 2,3, ...,k_2\} 
	\end{align}
\end{subequations}
where $Q_F, Q_G^{(j)}, Q_V^{(q)}\in\realv{n\times n}$ are symmetric matrices, $c_F$, $c_G^{(j)}$, $c_V^{(q)}$ $\in\realv{n}$ and $a_F$, $a_G^{(j)}$, $a_V^{(q)}$ $\in \realv{}$.

We start our consideration by introducing basic definitions and notations. Let the state space be $\mathcal{H}=\mathbb{C}^N$, with associated orthonormal basis $\left.\mathcal{E}=\{|j\rangle\right\}_{j \in\left\{0, \ldots, N-1\}\right.}$ and let the space of quantum states be denoted by $\mathcal{W}=\{\ket{\Psi} \in \mathbb{C}^N : \langle\Psi \mid \Psi\rangle=\| \ket{\Psi} \|^2=1 \}$. All operators will be represented on the $\mathcal{W}$ basis in the following. 
 
Considering a quantum system whose dynamics are governed by the controlled time-dependent Schr\"odinger equation 
 \begin{equation} \label{model1}
    i\hbar|\dot{\Psi}(t)\rangle=H(t)|\Psi(t)\rangle, \quad H(t)=H_c + \zeta(t) H_m,
\end{equation}
where we set $\hbar=1$, $\zeta(t)$ is the control input, $H_c$ is the drift (cost) Hamiltonian with corresponding eigenvalues $E_0,E_1,\dots,E_{N-1}$ and $H_m$ is the control (mixer) Hamiltonian. In this work, we consider both $H_c$ and $H_m$ to be time-independent, and they are non-commuting, i.e., $[H_c, H_m]\neq0$. 

The primary aim is to determine a feedback control law $\zeta(\ket{\Psi(t)})$ that ensures the convergence of the quantum system \eqref{model1} from any initial state to the ground state of the Hamiltonian $H_c$, denoted as $\ket{\Psi_g}=\text{argmin}_{\ket{\Psi}\in \mathcal{H}} \bra{\Psi}H_c\ket{\Psi}$. Consider a Lyapunov function in the form of $ V(\ket{\Psi}) = \bra{\Psi}H_c\ket{\Psi}$. The derivative of this Lyapunov function along the trajectories of system \eqref{model1} is given by $\dot{V}(\ket{\Psi(t)}) =  \bra{\Psi(t)}  \mathrm{i}[H_m,H_c] \ket{\Psi(t)} \zeta(t)$. Thus, designing $\zeta(t)$ as:
\begin{equation}
    \zeta(t)= -\bra{\Psi(t)}  i[H_m,H_c] \ket{\Psi(t)},
    \label{controller1}
\end{equation}
will guarantee $\dot{V}\leq0$. By applying the controller \eqref{controller1}, under certain assumptions (see Appendix A of \cite{magann2022lyapunov}), asymptotic convergence to the ground state $\ket{\Psi_g}$ is ensured. The time evolution operator associated to \eqref{model1} is given as:


\begin{align}
U(T, 0) = \eta e^{-i \int_{0}^{T} H(w) d w} \approx \prod_{k=1}^{d} e^{-i\left(\zeta(k \Delta t) H_{m} + H_{c}\right) \Delta t} \approx \prod_{k=1}^{d} e^{-i \zeta(k \Delta t)  H_{m}\Delta t}  e^{-iH_{c}\Delta t}
\end{align}
Here, $\eta$ is the time-ordering operator, $T=d \Delta t$ where $d$ is the number of piece-wise constant time intervals of length $\Delta t$, which is chosen to be small enough to maintain the approximation of $H(t)$ as a constant within each interval $\Delta t$, and in the last equation Trotter approximation is used. Hence, the evolved state becomes
\begin{align}
\ket{\Psi_d}    &= U(T,0) \ket{\Psi_0} =\prod_{k=1}^{d} \big( e^{-i \zeta(k \Delta t)  H_m\Delta t}e^{-iH_c\Delta t} \big) \ket{\Psi_0} =  \prod_{k=1}^{d} \big( M(\zeta_k)C \big)\ket{\Psi_0},
\label{evolution}
\end{align}
where we have introduced the notation $\zeta_k=\zeta(k \Delta t)$, $\ket{\Psi_k}=\ket{\Psi(k \Delta t)}$, $C=e^{-iH_c\Delta t}$, $M(\zeta_k)=e^{-i \zeta(k \Delta t)  H_m\Delta t}$. From \eqref{evolution}, it is seen that the values assigned to the controller at each discrete time step are equivalent to the circuit parameters. Thus, the terms \textit{controller} and \textit{circuit parameters} will be used interchangeably in this work.
For the control law, we adopt a discrete version of the controller \eqref{controller1} as follows:
\begin{equation}
    \zeta_{k+1}= - \bra{\Psi_k}  i[H_m,H_c]  \ket{\Psi_k} 
    \label{udis}
\end{equation}
By choosing $\Delta t$ to be sufficiently small, it can be guaranteed that the condition $\dot \zeta(t) \leq 0$ is satisfied (see Subsection 4-A of \cite{magann2022lyapunov} for details).  

For a drift Hamiltonian $H_c$, expressed as a summation of Pauli strings as $H_c=\sum_{r=1}^{m_0} a_r Q_r$, where $a_r$'s are real scalars, $m_0$ is given as a polynomial function of the number of qubits and $Q_r=Q_{r,1} \otimes Q_{r,2} \otimes ... \otimes Q_{r,n}$ with $Q_{r,q} \in \{ I,X,Y,Z \}$, the cost operator $C$ can be efficiently implemented as a quantum circuit (see \cite{magann2021digital} for details). Similarly, for efficient quantum circuit implementation of the mixer $M$, the design of the mixer Hamiltonian $H_m$ should have a similar structure as a sum of Pauli strings $ H_m = \sum_{r=1}^{m_1} b_r \bar Q_r $. For more insight on how to design the mixer Hamiltonian for FALQON, see \cite{malla2024feedback}.

FALQON implementation proceeds through the steps outlined below. 

\textbf{Step 1:} Initialize the algorithm by choosing a starting value for $\zeta_1=\zeta_\text{init}$, a time step $\Delta t$, and a maximum depth $d$ while setting $l=1$. In addition, design the cost Hamiltonian $H_m$.

\textbf{Step 2:} On the quantum computer, prepare the qubits into an easy-to-prepare initial state $\ket{\Psi_0}$. Subsequently, prepare the state $\ket{\Psi_l}$ by applying the quantum circuit $\ket{\Psi_l}=\prod_{k=1}^{l} \big( M(\zeta_k)C \big)\ket{\Psi_0}$.

\textbf{Step 3:} Compute the circuit parameter for the next layer $\zeta_{l+1}$. To estimate this parameter, expand it in terms of Pauli strings as follows:
\begin{equation}
    \zeta_{l+1}= -\bra{\Psi_l}  i[H_m,H_c]  \ket{\Psi_l} = \sum_{r=1}^{m_2} c_r \bra{\Psi_l} R_r \ket{\Psi_l},
\end{equation}
where $R_r$ is a Pauli string, and $m_2$ denotes the number of Pauli strings. Notably, $m_2$ depends on $m_0$ and $m_1$, both of which are polynomial functions of the qubit count. Consequently, $m_2$ is also a polynomial function of the qubit count.

\textbf{Step 4:} Add a new layer into the circuit by setting $l=l+1$,  and repeat Steps 2-4 iteratively until the maximum depth of layers is achieved $l=d$.

The resulting dynamically designed quantum circuit, $\prod_{k=1}^{d} \big( M(\zeta_k)C \big)$, alongside its parameters $\{\zeta_k\}_{k=1, \dots, d}$, represents the algorithm's output. This output effectively approximates the ground state of the Hamiltonian $H_c$.

\section{Feedback-Based Algorithm for Quantum Optimization with Constraints Based on Discrete-Time Lyapunov Control} \label{s3}
In this section, we propose a generalization of FALQON to tackle QCBO problems. To this aim, we design a new Hermitian operator $L$, which encodes the solution of the QCBO problem as its ground state. Subsequently, we devise a feedback control law to ensure that trajectories converge from any initial state to the ground state of this operator. To solve this control problem, we design a new feedback law to assign the circuit parameters. We call this new algorithm FALQON-C. In the following, we first introduce the discrete-time model for the evolution of the quantum circuit, and then describe FALQON-C. 

\subsection{Quantum Lyapunov Control for Discrete Systems} \label{ss31}
Instead of the continuous-time system, we consider the following discrete-time system:
\begin{equation} \label{model2}
    \ket{\Psi_{k+1}}=M(\zeta_k)C \ket{\Psi_k}
\end{equation}
Here, $k \in \mathbb{N}$ is the discrete-time index, $M(\zeta_k)=e^{-i\zeta_kH_m\Delta t}$ and $C=e^{-iH_c\Delta t}$. As seen in Section~2, this system results from a piece-wise constant discretization of the continuous-time system. The advantage of using the discrete-time system as opposed to the continuous-time system is that the former is directly realizable as a quantum circuit (see \cite{magann2021pulses} for further discussions).

We introduce a new Hermitian operator $L$ with eigenvalues $\omega_0, \omega_1, \dots, \omega_{N-1} $ such that it commutes with $H_c$, i.e., $[L,H_c]=0$. Defining the ground state of the operator $L$ to be $\ket{\Psi_f}=\text{argmin}_{\ket{\Psi}\in \mathcal{H}} \bra{\Psi} L \ket{\Psi}$, our main objective is to find a control law in a feedback form $\zeta(\bra{\Psi_k} A \ket{\Psi_k})$ where $A$ is an observable, that guarantees the convergence of the quantum system \eqref{model2} from any initial state to the desired final state, $\ket{\Psi_f}$. Note that restricting the controller to be in this form, i.e. $\zeta(\bra{\Psi_k} A \ket{\Psi_k})$, facilitates its evaluation using a quantum computer. Consider the Lyapunov function $ V(\ket{\Psi})=\bra{\Psi} L \ket{\Psi}$. In this case, we want to ensure that the Lyapunov function is non-increasing, i.e., $V(\ket{\Psi_{k+1}})-V(\ket{\Psi_k}) \leq 0$.

By first-order Taylor series expansion
\begin{equation} \label{T1}
    M(\zeta)=e^{-i\zeta H_m\Delta t} = I - i\Delta t H_m \zeta + O(\Delta t^2),
\end{equation}
\begin{equation} \label{T2}
    C=e^{-iH_c\Delta t} = I - i\Delta t H_c +  O(\Delta t^2),
\end{equation}
we have
\begin{align} \label{dV}
 V(\ket{\Psi_{k+1}})-V(\ket{\Psi_k})  &= \bra{\Psi_k} C M(\zeta_k) L M(\zeta_k) C \ket{\Psi_k} - \bra{\Psi_k} L \ket{\Psi_k} \nonumber \\
&= \zeta_k\Delta t \bra{\Psi_k}i[H_m,L]\ket{\Psi_k}+O(\Delta t^2).
\end{align}
To guarantee that $V(\ket{\Psi_{k+1}})-V(\ket{\Psi_k}) \leq 0$, we choose $\Delta t$ sufficiently small (see Remark~\ref{r1}), and design the feedback law as
\begin{equation} \label{controller2_dis}
    \zeta_{k+1}= -K   f( \Delta t \bra{\Psi_k}  i[H_m,L]  \ket{\Psi_k} ),
\end{equation}
where $K>0$ and $f(\cdot)$ is a continous function satisfying $xf(x)>0$ for all $x \neq 0$ and $f(0)=0$. Section~\ref{s4} will explore different design options for the function $f(\cdot)$. The application of the controller \eqref{controller2_dis} to the system \eqref{model2}, given certain assumptions are satisfied (see Section 2.1 of \cite{clausen2023measurement}), ensures asymptotic convergence to the desired state $\ket{\Psi_f}$ \cite{clausen2023measurement}. 
\begin{remark} \label{r1}
 Following the same analysis as in Section 4-A of \cite{magann2022lyapunov}, we obtain the following modified bound for $\Delta t$:
\begin{equation}
    |\Delta t|<\frac{\left|\bra{\Psi_k}  [H_m,L]  \ket{\Psi_k}\right|}{2\left(2 ||H_m|| \cdot||H_c||+\left|\bra{\Psi_k}  [H_m,L]  \ket{\Psi_k}\right|\right)\left(||H_c||+||H_m|| \left| \zeta_k \right|\right)}
\end{equation}
Choosing $\Delta t$ according to this bound guarantees that the Lyapunov function is non-increasing, i.e. $V(\ket{\Psi_{k+1}})-V(\ket{\Psi_k}) \leq 0$.
\end{remark}

\begin{remark} \label{r2}
    In case of multiple controllers (see Subsection 2-C of \cite{magann2022lyapunov} for details), the mixer becomes $M(\zeta_k)=e^{-\sum_{p=1}^{P}i\zeta_k^{(p)} H_m^{(p)}\Delta t}$, where $P$ is the number of control inputs, while the controller \eqref{controller2_dis} is modified to $\zeta^{(p)}_{k+1}= -K_p   f( \Delta t \bra{\Psi_k}  i[H_m^{(p)},L]  \ket{\Psi_k} )$.
\end{remark}

\subsection{Feedback-Based Algorithm for Quantum Optimization with Constraints} \label{ss32}

We begin by providing a detailed explanation of the construction process for the operator $L$. Following this, we detail the algorithmic steps of FALQON-C. Given the problem defined as \eqref{QCBO}, we need to construct the operator $L$ such that its ground state encodes the solution to the problem. The design principles ensuring convergence to the ground state of the operator $L$ are outlined as follows \cite{clausen2023measurement}:
\begin{itemize}
    \item  The operator $L$ must commute with $H_c$, i.e. $[L,H_c]=0$.
    \item All eigenvalues of $L$ must be distinct denoted as $\omega_l \neq \omega_k$ for all $l \neq k$.
    \item The eigenvalue $\omega_f$ corresponding to the target eigenstate should be the minimum among all eigenvalues, indicated as $\omega_f < \omega_k$ for $k \in (1,2,...,N-2)$ where $\omega_k \neq \omega_f$.
\end{itemize}
We start by converting the cost function $F$ in \eqref{QCBO} into a cost Hamiltonian $H_c$. For the constraints, the inequality constraints $G^{(j)}$ should be converted firstly into equality constraints, then converted alongside the equality constraints $V^{(q)}$ into penalty Hamiltonians ${H_p^{(j)}}$, $j\in \{1,2,\dots,k_3:=k_1+k_2\}$. This conversion involves mapping each binary variable $y_j$ to a Pauli-Z operator using the transformation $y_j \mapsto \frac{1}{2}(I-Z_j)$, where $Z_j$ is the Pauli Z operator applied to the $j^{th}$ qubit, such that the resulting Hamiltonians satisfy $H_c\ket{y}=F(y)\ket{y}$ and $H_p^{(j)}\ket{y}=W^{(j)}(y) \ket{y}$ \cite{hadfield2021representation}, where the penalty function $W^j(y)$ is defined as  $ W^{(j)}(y)=|G^{(j)}(y)|^a$. Note that we need $W^{(j)}(y)>0$ $\forall y$ to ensure that each eigenvalue in the operator $L$ associated with an infeasible outcome is shifted positively, i.e. have larger values. In our case, we choose a=2, hence we get $W^{(j)}(y)=(G^{(j)}(y))^2$. Define the operator:
\begin{equation}
    L := H_c+\sum_{j=1}^{k_3} \gamma_j H_p^{(j)}
    \label{LyaOp}
\end{equation}
where $\{\gamma_j\}_{j=1, \dots, k_3}$ are hyperparameters that should be chosen large enough to guarantee that the ground state of the operator $L$ encodes the optimal feasible solution. Defining the set of feasible outcomes as $\mathcal{Y}_\text{f} = \{y \in \{0,1\}^n : W^{(j)}(y)=0, \; j\in \{1,2, \dots, k_3 \} \}$  and the set of infeasible outcomes as $\mathcal{Y}_\text{inf}=\{y \in \{0,1\}^n : W^{(j)}(y) \neq 0, \; j\in \{1,2, \dots, k_3 \} \}$, then the eigenvalues of the operator $L$ are given as:
\begin{equation}
\omega_q(L)= 
\begin{cases}
   E_q  & \text{for } q \in \mathcal{Y}_\text{f} \\
   E_q+ \sum_j \gamma_j W^j(q)  & \text{for } q \in \mathcal{Y}_\text{inf}
\end{cases}
\end{equation}
while $L$ shares the same set of eigenvectors as $H_c$, satisfying $[H_c,L]=0$.

Therefore, to guarantee that the smallest eigenvalue of $L$ corresponds to an eigenvector that encodes the solution to the problem \eqref{QCBO}, we need to choose $\gamma_j$ to satisfy the following condition: 
\begin{align}
     \sum_j \gamma_j W^{(j)}(q) & \geq E^\text f_{\min} - E_{\min}, \quad q\in \mathcal{Y}_\text{inf},
\end{align}
where $E_{\min}$ is the minimum eigenvalue of $H_c$ and $E^f_{\min}$ is the smallest eigenvalue of $H_c$ that corresponds to an eigenvector encoding a feasible outcome. Define the energy gap $E_g := E_{\max} - E_{\min} > E^f_{\min} - E_{\min}$ where $E_{\max}$ is the largest eigenvalue of $H_c$. For a cost Hamiltonian expanded in the Pauli basis as $H_c=\sum_{r=1}^{m_0} c_r Q_r$, where $Q_r$ are Pauli strings then the upper bound is $E_g \leq 2||H_c|| \leq 2 \sum_r \abs{c_r} $. We assume that, without loss of generality, for each $q \in \mathcal{Y}_\text{inf}$, the value of $W^{(j)}(q)$ is larger than or equal to $1$ at least for one $j$, which can be guaranteed by multiplying the equality constraint by the least common multiple of the denominators of the coefficients. Choosing each of the shifting parameters $\gamma_j$ to be larger than $E_g$, we get
\begin{align}
     \sum_j \gamma_j W^{(j)}(q) & \geq E_g \sum_j W^{(j)}(q)  \geq E_g > E^\text f_{\min} - E_{\min}, \quad q\in \mathcal{Y}_\text{inf},
\end{align}
which guarantees shifting the eigenvalues corresponding to eigenvectors that encode infeasible outcomes to be larger than the ones encoding feasible outcomes. Simulation results indicate that opting for such a choice is unnecessarily large, and better performance could be achieved by choosing smaller values, provided they are sufficiently large to ensure that the ground state of $L$ corresponds to the minimum energy feasible state. We now introduce FALQON-C as follows:


\textbf{Step 1:} Initialize the algorithm by choosing a starting value for $\zeta_1=\zeta_\text{init}$, a time step $\Delta t$, a maximum depth $d$, the shifting hyperparameters $\{\gamma_j\}_{j=1, \dots, k_3}$, while setting $l=1$. Additionally, design the cost Hamiltonian $H_m$ and the operator $L$ using \eqref{LyaOp}.

\textbf{Step 2:} On the quantum computer, prepare the qubits into an easy-to-prepare initial state $\ket{\Psi_0}$. Subsequently, prepare the state $\ket{\Psi_l}$ by applying the quantum circuit $\ket{\Psi_l}=\prod_{k=1}^{l}\big( M(\zeta_k)C\big) \ket{\Psi_0} $.

\textbf{Step 3:} Compute the circuit parameter for the next layer $\zeta_{l+1}$ using \eqref{controller2_dis}. Since the operator $L$ is given in terms of Pauli strings, then the controller \eqref{controller2_dis} can be expanded in terms of Pauli strings as $  \zeta_{l+1}= -f(\bra{\Psi_l}  \mathrm{i}[H_m,L]  \ket{\Psi_l}) = f(\sum_{r=1}^{m_3} d_r\bra{\Psi_l} S_r \ket{\Psi_l}$), where $S_r$ is a Pauli string.

\textbf{Step 4:} Add a new layer into the quantum circuit by setting $l=l+1$,  and repeat Steps 2-4 iteratively till the maximum depth is reached $l=d$.

The resulting output effectively approximates the ground state of the operator $L$, which encodes the solution to the QCBO problem.

\section{Application to a QCBO Problem} \label{s4}
To evaluate the performance of the proposed algorithm, we use two performance metrics, namely the approximation ratio and the success probability. For a given state $\ket{\Psi}$, the approximation ratio $r_a$ is defined as $ r_a= \frac{V(\ket{\Psi})-\omega_{\max}}{\omega_{\min}-\omega_{\max}}$, where $\omega_{\max}$ and $\omega_{\min}$ are the maximum and minimum eigenvalues of $L$, respectively \cite{herman2023constrained}. The approximation ratio is an essential metric since it evaluates the closeness of the Lyapunov function $V(\ket{\Psi})$ to the optimal solution of the optimization problem. The success probability is defined as $ P_s\equiv \bigl|\bra{y^*}\ket{\Psi}\bigr|^2$, where the bit string $y^*$ is defined as the optimal solution to the optimization problem.

Consider the following QCBO problem:
	\begin{align*} \label{toy1} 
		&\min _{x \in \{0,1\}^{3}} F(x)= -2x_1-5x_2-3x_3-2x_1x_2 \nonumber\\
		&\text{s.t.} \quad G(x) = 1-x_1 -3 x_2 -x_3 = 0    
  \end{align*}
From $G(x)$ , we observe that the feasible set of states is $B_1=\{\ket{100}, \ket{001}\}$, while the infeasible set of states is $B_2=\{\ket{000}, \ket{010},\ket{011},\ket{101},\ket{110},\ket{111}\}$. 

To apply FALQON-C, we first construct the cost Hamiltonian by transforming $F(x)$ into a Hamiltonian to get $H_c=1.5Z_1 +3Z_2 +1.5Z_3 -0.5Z_1Z_2 -5.5I=\text{diag}(0,-3,-5,-8,-2,-5,-9,-12)$. Similarily, we construct the penalty Hamiltonian from $W^{(1)}(x) = (G(x))^2=(1 -{x}_1 -3 {x}_2 - {x}_3)^2=1-{x}_1+3{x}_2-{x}_3+6{x}_1{x}_2+2{x}_1{x}_3+6{x}_2{x}_3$, where we used ${x}_i^2={x}_i$, to get $H_p^{(1)}=-1.5Z_1 -4.5Z_2 -1.5Z_3 +1.5Z_1Z_2 +0.5Z_1Z_3 +1.5Z_2Z_3 + 5I=\text{diag}(1,0,4,9,0,1,9,16)$. We then construct the operator $L=H_c+3 H_p^{(1)} = -3Z_1 -10.5Z_2 - 3Z_3 +4Z_1Z_2 +1.5 Z_1Z_3 +4.5Z_2Z_3 + 9.5I = \text{diag}(3,-3,7,19,-2,-2,18,36) $ where we chose $\gamma_1=3$. It is clear that the operator $L$ encodes the minimum energy feasible state $\ket{001}$ corresponding to eigenvalue $-3$ as its ground state. 


To solve the problem using FALQON, we first convert the problem into an equivalent QUBO problem. We do so by incorporating the equality constraints into the cost function as a penalty term to get the following QUBO problem, $\min _{x \in \{0,1\}^{3}}  F(x)+3 (G(x))^2 = -6x_1-3x_3+6x_1x_2 + 2x_1x_3+4x_2x_3 $. We map this into the following cost Hamiltonian: $\bar H_c=-3Z_1 -10.5Z_2 - 3Z_3 +4Z_1Z_2 +1.5 Z_1Z_3 +4.5Z_2Z_3 + 9.5I = \text{diag}(3,-3,7,19,-2,-2,18,36)$. The control Hamiltonian is designed as $H_m=\sum_{p=1}^3 \zeta^{(p)} H_m^{(p)}=\sum_{p=1}^3 \zeta^{(p)} X_p $. Figure 1 shows the quantum circuit of the cost operator for FALQON and FALQON-C. This figure shows that FALQON results in a deeper circuit compared to FALQON-C due to the terms added from the constraints. These terms usually include $Z_iZ_j$ terms that need control gates to be implemented as a quantum circuit.  Note that $\bar H_c$ and $L$ are equivalent. In FALQON, this operator is used to implement the quantum circuit and to evaluate the controller, while in FALQON-C, the operator $L$ is only used to evaluate the controller. 
\begin{figure}[H]
    \centering
    \begin{subfigure}[t]{0.50\textwidth}
        \centering
        \includegraphics[width=0.48\linewidth]{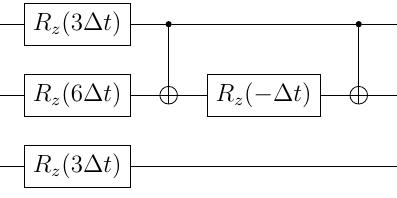}
    \end{subfigure}%
    ~ 
    \begin{subfigure}[t]{0.5\textwidth}
        \centering
        \includegraphics[width=\textwidth]{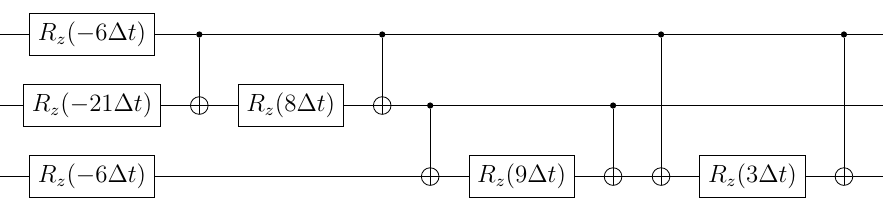}
    \end{subfigure}
    \caption{The quantum circuit for the cost operator of FALQON-C $C=e^{-iH_c\Delta t}$ (to the left) and FALQON $C=e^{-i\bar H_c \Delta t}$(to the right). }
\end{figure}
We run simulations for FALQON and FALQON-C for a depth of $200$ layers using the statevector simulator. The initial state is chosen as the equal superposition state $\ket{\Psi_0}=\ket{+}^{\otimes 3}$. The time step is set to be $\Delta t = 0.02$, the controller's gains are chosen to be $K_1=K_2=K_3=1$, and the initial guess for the controllers is $0$. The results are given in Figure 2. Figure 2 shows that FALQON-C achieves better performance compared to FALQON. 

\begin{figure}[H]
    \centering
    \begin{subfigure}[t]{0.45\textwidth}
        \centering
        \includegraphics[width=\textwidth]{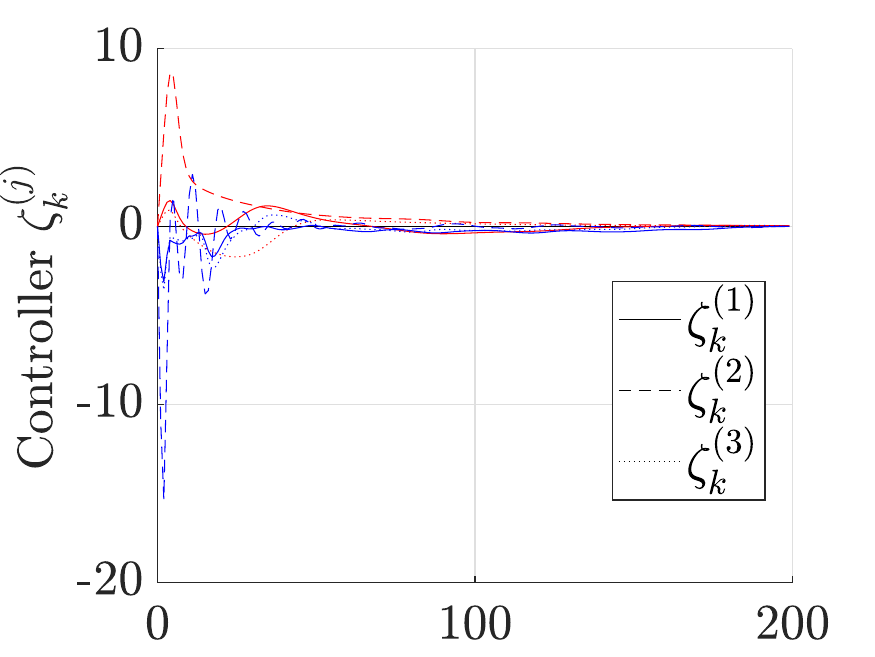}
    \end{subfigure}
    \begin{subfigure}[t]{0.45\textwidth}
        \centering
        \includegraphics[width=\textwidth]{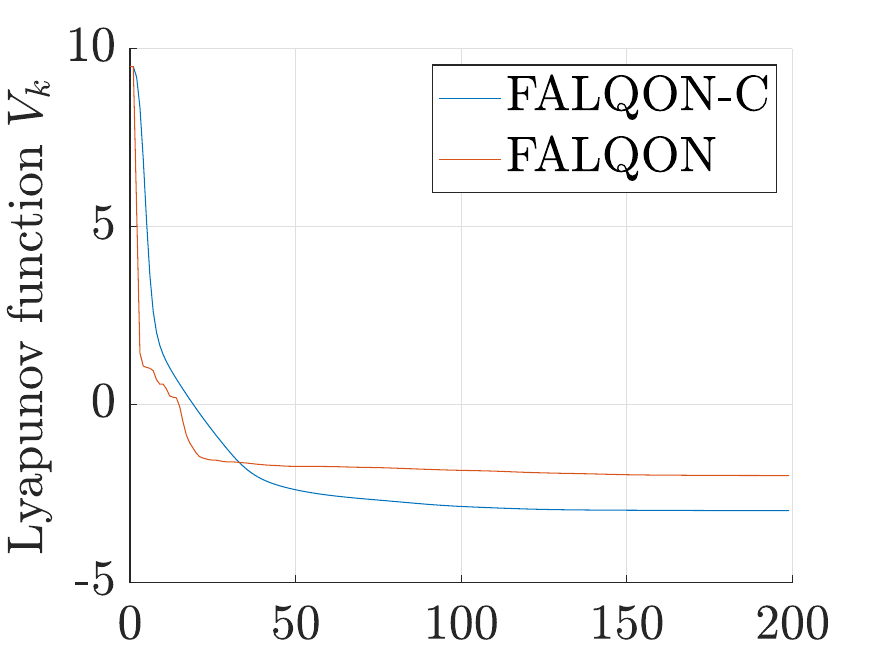}
    \end{subfigure}
    \\
    \begin{subfigure}[t]{0.43\textwidth}
        \centering
        \includegraphics[width=\textwidth]{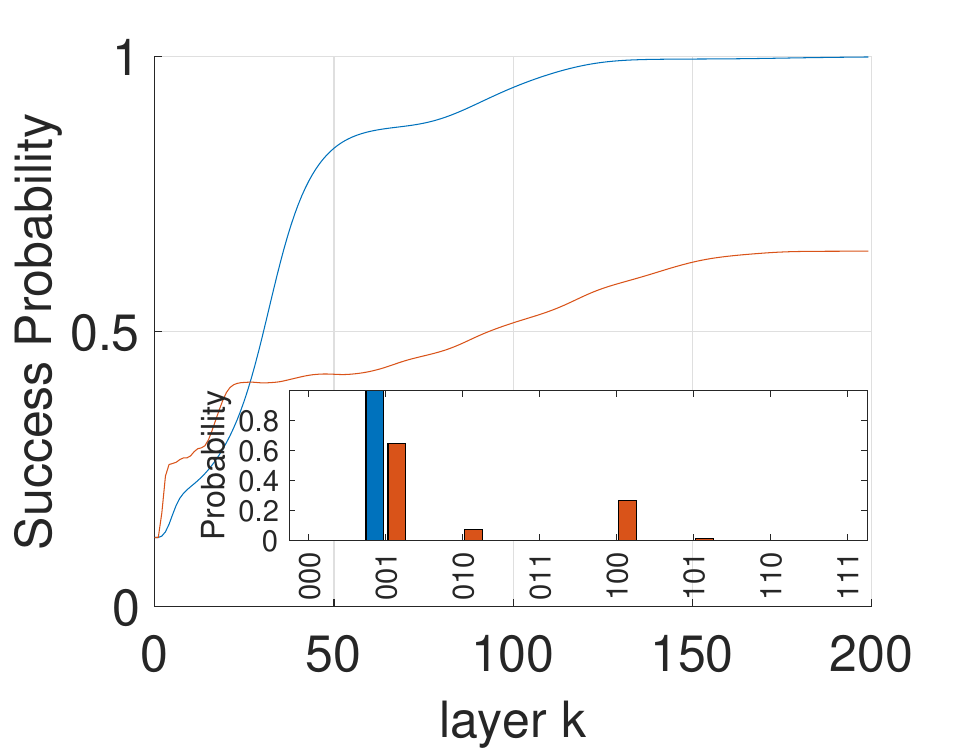}
    \end{subfigure}
    \begin{subfigure}[t]{0.45\textwidth}
        \centering
        \includegraphics[width=\textwidth]{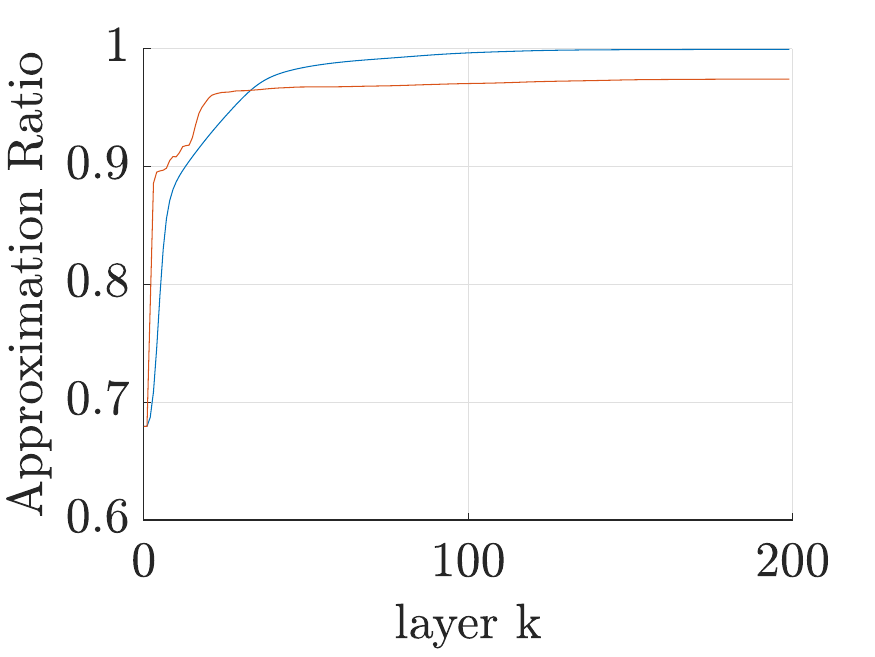}
    \end{subfigure}
    \caption{Comparison between the simulation results for running FALQON-C and FALQON to solve the QCBO problem. The layer index $k$ is plotted versus: (a) The controllers, (b) The Lyapunov function, (c) The success probability (a histogram of the final state) and (d) The approximation ratio.}
\end{figure}  

In literature, several methods are proposed for choosing the function $f(\cdot)$ in the design of the controller \eqref{controller2_dis}. These include the following:
\begin{itemize}
    \item Standard Lyapunov Controller \cite{magann2022lyapunov}: $f_\text{standard}(\ket{\Psi_k})= \bra{\Psi_k}  i[H_m,L]  \ket{\Psi_k}$.
    \item Bang-bang Controller \cite{kuang2017rapid}: $f_\text{bang-bang}(\ket{\Psi_k})= \text{sign}(\bra{\Psi_k}  i[H_m,L]  \ket{\Psi_k})$.
     \item Finite-Time Lyapunov Controller 1 \cite{kuang2021finite}: \\$f_\text{finite-1}(\ket{\Psi_k})=  \bra{\Psi_k}  i[H_m,L]  \ket{\Psi_k} |\bra{\Psi_k}  i[H_m,L]  \ket{\Psi_k}|^{c_1}$.
     \item Finite-Time Lyapunov Controller 2 \cite{kuang2021finite}: \\$f_\text{finite-2}(\ket{\Psi_k})=  \text{sign}(\bra{\Psi_k}  i[H_m,L]  \ket{\Psi_k}) |\bra{\Psi_k}  i[H_m,L]  \ket{\Psi_k}|^{c_1}$.
    \item Fixed-Time Lyapunov Controller \cite{li2022lyapunov}: \\
        $ f_\text{fixed}(\ket{\Psi_k})=  -K_1\text{sign}(\bra{\Psi_k}  i[H_m,L]  \ket{\Psi_k}) |\bra{\Psi_k}  i[H_m,L]  \ket{\Psi_k}|^{c_1}  \\
          - K_2\text{sign}(\bra{\Psi_k}  i[H_m,L]  \ket{\Psi_k}) |\bra{\Psi_k}  i[H_m,L]  \ket{\Psi_k}|^{c_2}  $.
\end{itemize}
where $c_1 \in (0,1)$ and $c_2=1/c_1$. We run FALQON-C for these controllers and compare their performance where the controllers' parameters are chosen to be $c_1=0.90$, $c_2=1/c_1$, $K=1$, $K_1=K_2=1$ and for the bang-bang controller $K=3.5$. The results are shown in Figure 3. From Figure 3, we see that the fixed-time controller 
gives a slightly better performance than the other designs in terms of convergence speed, final approximation ratio, and success probability. It is also observed that the bang-bang controller achieves faster convergence; however, it converges to lower values of $SP \approx 0.95$ and $r_a \approx 0.99$.

   \begin{figure}[H] \label{2}
    \centering
    \begin{subfigure}[t]{0.31\textwidth}
        \centering
        \includegraphics[width=\textwidth]{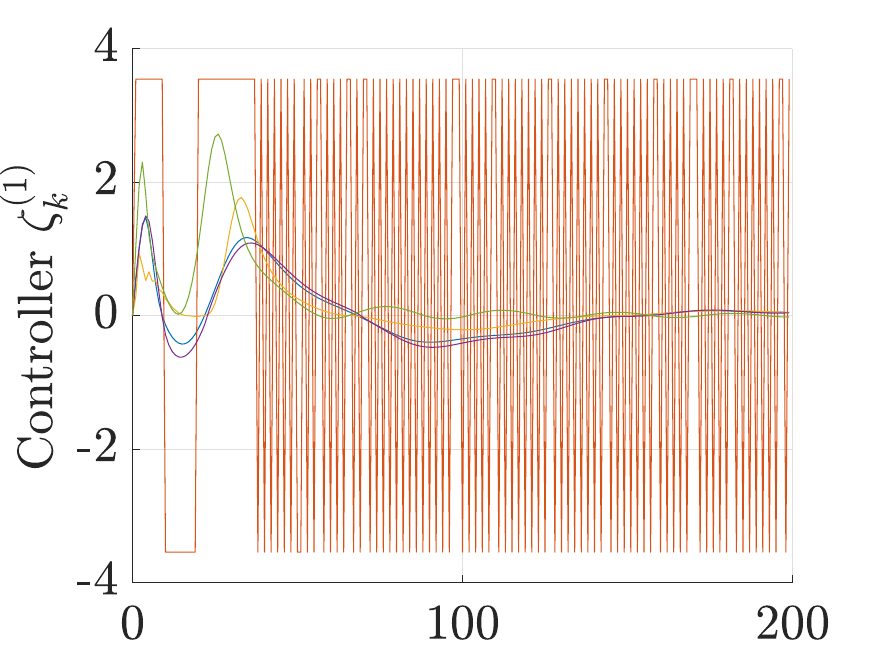}
    \end{subfigure}%
    ~ 
    \begin{subfigure}[t]{0.31\textwidth}
        \centering
        \includegraphics[width=\textwidth]{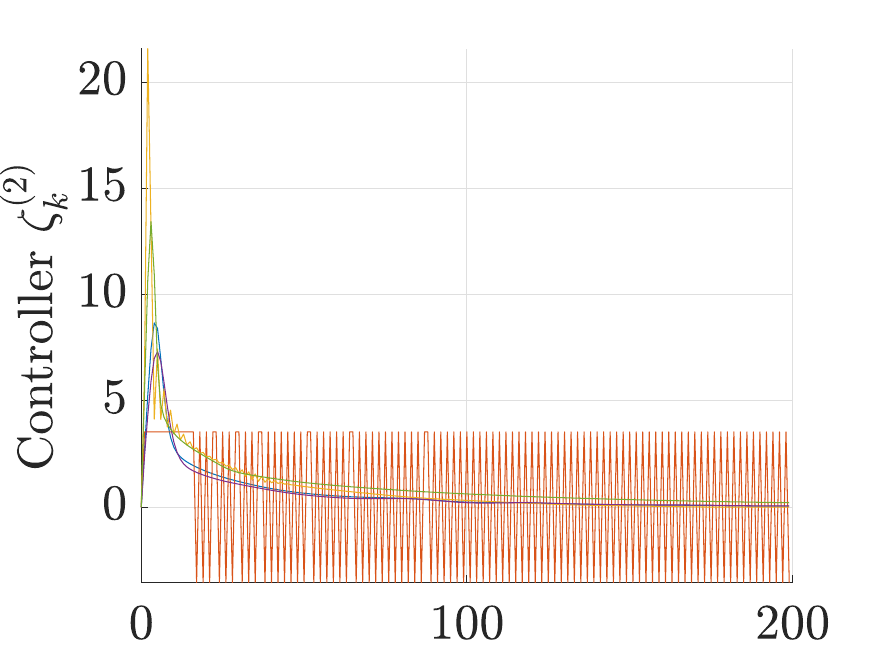}
    \end{subfigure}
    ~
    \begin{subfigure}[t]{0.31\textwidth}
        \centering
        \includegraphics[width=\textwidth]{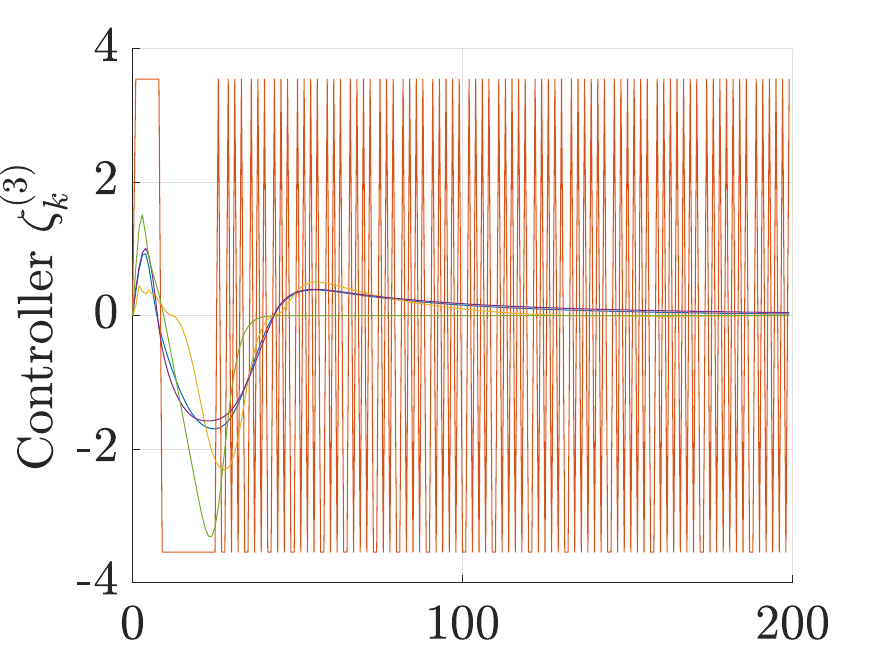}
    \end{subfigure}

    \vspace{0.5cm} 

    \begin{subfigure}[t]{0.31\textwidth}
        \centering
        \includegraphics[width=\textwidth]{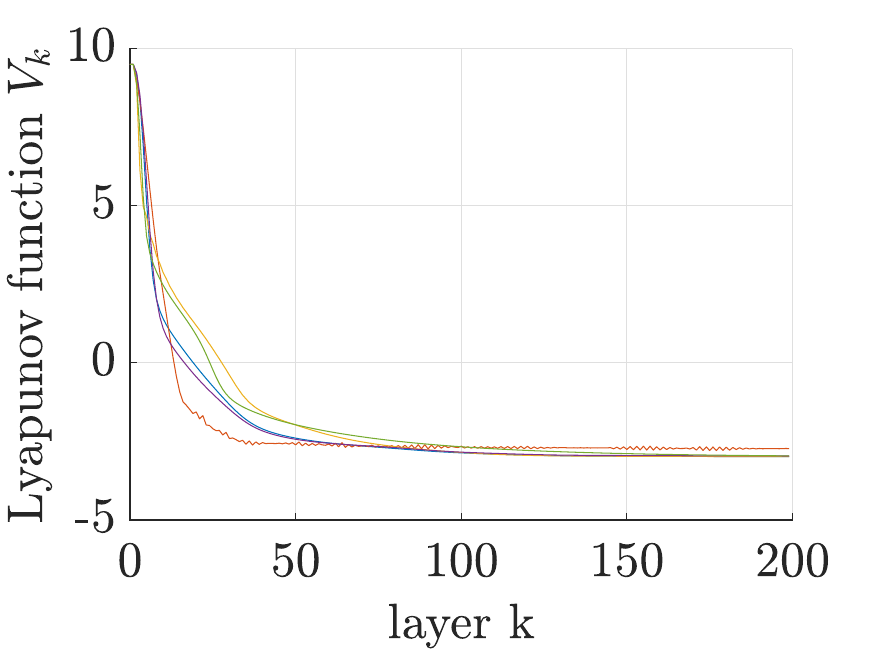}
    \end{subfigure}%
    ~ 
    \begin{subfigure}[t]{0.31\textwidth}
        \centering
        \includegraphics[width=\textwidth]{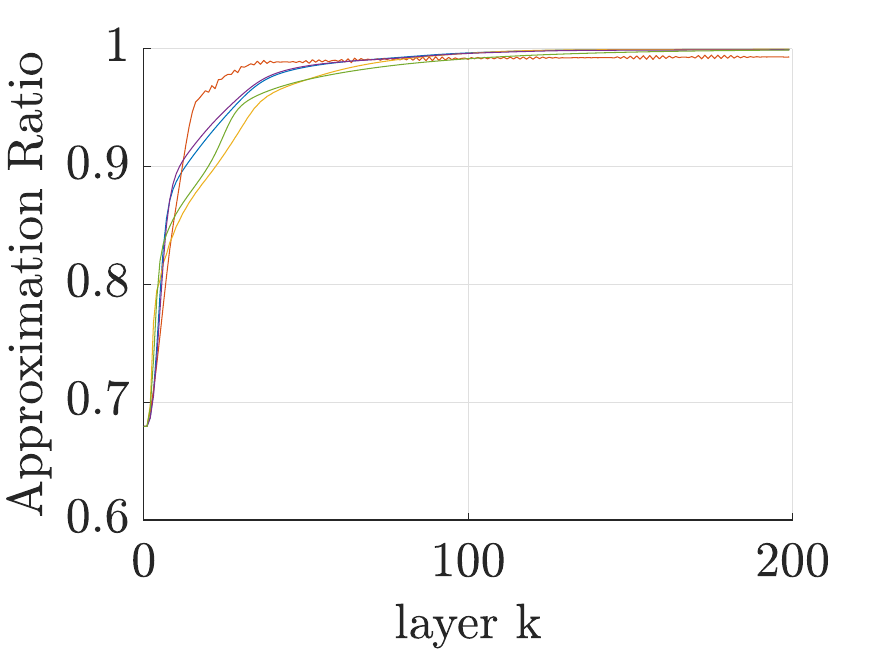}
    \end{subfigure}
    ~
    \begin{subfigure}[t]{0.31\textwidth}
        \centering
        \includegraphics[width=\textwidth]{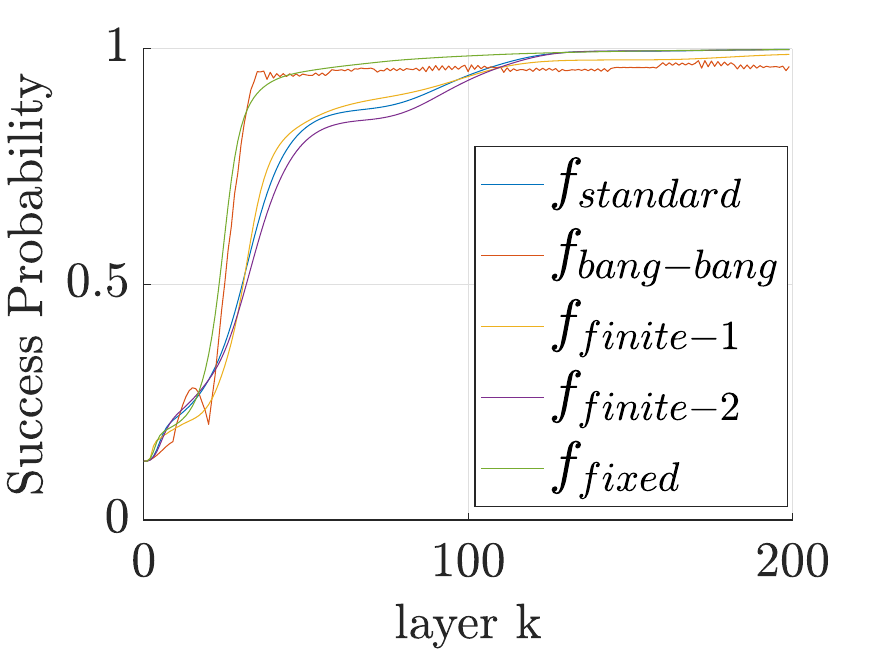}
    \end{subfigure}
    
    \caption{Simulation results of FALQON-C for the different controllers.}
\end{figure}


\section{Conclusion and Future Work} \label{s6}
We introduced FALQON-C as an enhancement of FALQON to solve QCBO problems more efficiently. Compared to FALQON, our algorithm reduces the quantum resources by decreasing the depth of the circuit and can achieve better performance. We showed the effectiveness of our proposed algorithm for solving QCBO problems. 

In future work, several avenues exist for enhancing the efficacy of feedback-based quantum algorithms. 
One potential avenue is exploring alternative ways to the Trotterized evolution in the design of the algorithm, such as the linear combination of unitaries and quantum signal processing algorithms \cite{chen2021quantum}. Additionally, while our work employs the equal superposition state as the initial state, integrating warm-starting techniques could reduce circuit depth and enhance efficiency. 
\newpage
This work extends the application of feedback-based quantum algorithms to a more general class of complex engineering problems and provides valuable insights into the relation between control theory and quantum algorithms for optimization. \\ \\
\textbf{Acknowledgment.} This work was supported by Independent Research Fund Denmark (DFF), project number 0136-00204B.

%
%

 \bibliographystyle{splncs04}
 \bibliography{mybibliography}

\begin{thebibliography}{10}
\providecommand{\url}[1]{\texttt{#1}}
\providecommand{\urlprefix}{URL }
\providecommand{\doi}[1]{https://doi.org/#1}

\bibitem{bharti2022noisy}
Bharti, K., Cervera-Lierta, A., Kyaw, T.H., Haug, T., Alperin-Lea, S., Anand, A., Degroote, M., Heimonen, H., Kottmann, J.S., Menke, T., et~al.: Noisy intermediate-scale quantum algorithms. Reviews of Modern Physics  \textbf{94}(1),  015004 (2022)

\bibitem{cerezo2021variational}
Cerezo, M., Arrasmith, A., Babbush, R., Benjamin, S.C., Endo, S., Fujii, K., McClean, J.R., Mitarai, K., Yuan, X., Cincio, L., et~al.: Variational quantum algorithms. Nature Reviews Physics  \textbf{3}(9),  625--644 (2021)

\bibitem{chen2021quantum}
Chen, Y.H., Kalev, A., Hen, I.: Quantum algorithm for time-dependent hamiltonian simulation by permutation expansion. PRX Quantum  \textbf{2}(3),  030342 (2021)

\bibitem{clausen2023measurement}
Clausen, H.G., Rahman, S.A., Karabacak, {\"O}., Wisniewski, R.: Measurement-based control for minimizing energy functions in quantum systems. IFAC-PapersOnLine  \textbf{56}(2),  5171--5178 (2023)

\bibitem{fernandez2022study}
Fern{\'a}ndez-Pend{\'a}s, M., et~al.: A study of the performance of classical minimizers in the quantum approximate optimization algorithm. Journal of Computational and Applied Mathematics  \textbf{404},  113388 (2022)

\bibitem{grimsley2019adaptive}
Grimsley, H.R., Economou, S.E., Barnes, E., Mayhall, N.J.: An adaptive variational algorithm for exact molecular simulations on a quantum computer. Nature communications  \textbf{10}(1), ~3007 (2019)

\bibitem{hadfield2021representation}
Hadfield, S.: On the representation of boolean and real functions as hamiltonians for quantum computing. ACM Transactions on Quantum Computing  \textbf{2}(4),  1--21 (2021)

\bibitem{herman2023constrained}
Herman, D., Shaydulin, R., Sun, Y., Chakrabarti, S., Hu, S., Minssen, P., Rattew, A., Yalovetzky, R., Pistoia, M.: Constrained optimization via quantum zeno dynamics. Communications Physics  \textbf{6}(1), ~219 (2023)

\bibitem{kuang2017rapid}
Kuang, S., Dong, D., Petersen, I.R.: Rapid lyapunov control of finite-dimensional quantum systems. Automatica  \textbf{81},  164--175 (2017)

\bibitem{kuang2021finite}
Kuang, S., Guan, X., Dong, D.: Finite-time stabilization control of quantum systems. Automatica  \textbf{123},  109327 (2021)

\bibitem{larsen2023feedback}
Larsen, J.B., Grace, M.D., Baczewski, A.D., Magann, A.B.: Feedback-based quantum algorithm for ground state preparation of the fermi-hubbard model. arXiv preprint arXiv:2303.02917  (2023)

\bibitem{li2022lyapunov}
Li, X., Wen, C., Wang, J.: Lyapunov-based fixed-time stabilization control of quantum systems. Journal of Automation and Intelligence  \textbf{1}(1),  100005 (2022)

\bibitem{magann2021pulses}
Magann, A.B., Arenz, C., Grace, M.D., Ho, T.S., Kosut, R.L., McClean, J.R., Rabitz, H.A., Sarovar, M.: From pulses to circuits and back again: A quantum optimal control perspective on variational quantum algorithms. PRX Quantum  \textbf{2}(1),  010101 (2021)

\bibitem{magann2021digital}
Magann, A.B., Grace, M.D., Rabitz, H.A., Sarovar, M.: Digital quantum simulation of molecular dynamics and control. Physical Review Research  \textbf{3}(2),  023165 (2021)

\bibitem{magann2022feedback}
Magann, A.B., Rudinger, K.M., Grace, M.D., Sarovar, M.: Feedback-based quantum optimization. Physical Review Letters  \textbf{129}(25),  250502 (2022)

\bibitem{magann2022lyapunov}
Magann, A.B., Rudinger, K.M., Grace, M.D., Sarovar, M.: Lyapunov-control-inspired strategies for quantum combinatorial optimization. Physical Review A  \textbf{106}(6),  062414 (2022)

\bibitem{malla2024feedback}
Malla, R.K., Sukeno, H., Yu, H., Wei, T.C., Weichselbaum, A., Konik, R.M.: Feedback-based quantum algorithm inspired by counterdiabatic driving. arXiv preprint arXiv:2401.15303  (2024)

\bibitem{rahman2024feedback}
Rahman, S.A., Karabacak, {\"O}., Wisniewski, R.: Feedback-based quantum algorithm for excited states calculation. arXiv preprint arXiv:2404.04620  (2024)

\bibitem{rahman2024weighted}
Rahman, S.A., Karabacak, {\"O}., Wisniewski, R.: Weighted feedback-based quantum algorithm for excited states calculation. arXiv preprint arXiv:2404.19386  (2024)

\bibitem{DavidWakeham2021}
Wakeham, D., Ceroni", J.: {"Feedback-Based Quantum Optimization (FALQON)"} ("5" "2021"), \url{https://pennylane.ai/qml/demos/tutorial\_falqon/}, date: 2024-02-26

\end{thebibliography}

\end{document}